**Observation and Interpretation of Field Emission Saturation Induced by an Ultrafast Intense Terahertz Field**

Wentao Yu[1,2], Nongchao Tan[1,2,3*], Kai Peng[1,2], Kai Jiang[1,2], Zhao Yun[1,2], Sijie Fan[1,2], Longding Wang[1,2], Yixiao Fu[1,2], Renkai Li[1,2], Yingchao Du[1,2], Lixin Yan[1,2], Chuanxiang Tang[1,2], Wenhui Huang[1,2*]

[1]Key Laboratory of Particle and Radiation Imaging, Tsinghua University, Ministry of Education, Beijing 100084, China

[2]Department of Engineering Physics, Tsinghua University, Beijing 100084, China

[3]Department of Nuclear Science and Technology, College of Science, National University of Defense Technology, Changsha 410073, China

**\*Corresponding authors**

Nongchao Tan, e-mail: tannongchao@nudt.edu.cn, tel: +8615573160675;

Wenhui Huang, e-mail: huangwh@mail.tsinghua.edu.cn, tel: +8613661391696.

**Abstract**

Field emission under ultrafast intense terahertz (THz) fields provides a promising approach for generating electron bunches with ultrashort pulse durations and high charge densities. It is generally believed that the field emission current described by Schottky–Nordheim theory increases dramatically with the applied electric field. However, we conducted extensive field emission experiments using quasi-single-cycle strong-field THz radiation at various energy levels and different temperatures and observed an intriguing phenomenon where the emitted charge reached saturation. A novel model is proposed to interpret this phenomenon, which considers the contribution of surface valence electrons and the dynamic replenishment of free electrons from the bulk to the surface. The experimentally observed convex relationship between the emitted charge and THz energy is consistent with the model's prediction, unlike the concave relationship derived from the traditional Schottky–Nordheim formula. In addition, another observed counterintuitive phenomenon, the inverse correlation between the cathode temperature and saturated emission charge, is also well interpreted by the model. This work offers comprehensive insights into field emission dynamics under ultrafast intense fields, paving the way for generating electron bunches with unprecedented temporal resolution.

Ultrafast processes have attracted considerable global interest because of their inherent fascination and immense potential in contemporary scientific research[1-6]. These processes, such as transient dynamics in chemical reactions and rapid structural transformations after laser pulse pumping, primarily unfold across femtosecond to picosecond timescales and are highly important for scientific understanding and technological innovation. To investigate these ultrafast phenomena in depth, researchers require advanced probes with sufficient temporal and spatial resolving capacity to decode the spatiotemporal complexity. The commonly used solution employs an ultrafast high-brightness electron beam or electron-derived X-rays[7-11]. Such electron sources must be able to generate beams exhibiting ultrashort temporal structures, extremely low emittances, and narrow energy spreads, thereby ensuring sufficient resolution for capturing ultrafast transformations with temporal precision[12]. This imperative imposes escalating requirements on high-quality electron production, becoming a critical research frontier and a persistent technical challenge in beam physics[13,14].

Field emission electron sources, which can generate coherent electron bunches with ultrahigh charge densities in the nanoscale emission area, have garnered increasing attention and achieved tremendous development[15-18]. Field emission dynamics rely on substantial field intensification derived from metal tips or other nanoscale structures and exhibit nearly instantaneous responses to field variation. Hence, it has evolved into a cornerstone methodology for producing ultrafast, high-brightness electron beams[19-28]. These electron sources exploit light–matter interactions under ultrafast laser or terahertz (THz) excitation to generate electron beams with extremely short pulse durations, providing unprecedented temporal resolution for ultrafast dynamic studies. Therefore, investigating the emission characteristics of ultrashort and extremely intense fields is crucial for high-quality electron generation. The existing field emission paradigms, epitomized by the Fowler–Nordheim or Schottky–Nordheim (S–N) theory, are typically derived in the DC or low-frequency regimes[29,30]. Although these models have demonstrated remarkable success in explaining conventional field emission phenomena[31,32], their applicability with ultrafast, extremely intense and high-frequency electromagnetic excitations needs further attention and analysis[33,34]. For example, extremely high current density emission under extremely strong field strengths can only be studied via ultrafast intense electromagnetic pulses because the emitter melts at such high emission current density under DC or low-frequency conditions.

In this study, we developed a THz-driven electron gun to investigate the ultrafast field emission process. This gun is composed of a tapered rectangular waveguide and a nanotip emitter to achieve considerable field enhancement. By applying quasi-single-cycle THz fields with a pulse duration of several picoseconds and peak field strengths exceeding 16 GV/m, emission charge saturation has been experimentally observed from the metal tip, which is distinct from the prediction of the S–N theory. To interpret this phenomenon, we propose a novel model that comprehensively accounts for quantum tunnelling effects under intense fields, surface valence electron (SVE) density, and the free-electron replenishment (FER) rate inside a metal. The total field emission charge, resulting from the charge associated with the SVEs and the contribution of FER currents, matches the experimentally measured saturated emission charge. We also validated our concept by investigating the influence of the cathode temperature on the saturated emission charge. The established model enables a more precise description and prediction of ultrafast intense field emission behaviour.

**Concept and implementation**

The experimental setup schematic is shown in Fig. 1. The quasi-single-cycle THz pulse (see Method) is coupled into a well-designed electron gun fabricated from oxygen-free copper (see Method), with the electric field aligned with electron propagation. This gun features a tapered rectangular waveguide horn structure for improving the coupling efficiency and field enhancement. The terminal short-circuit boundary serves as a mirror to reflect the THz pulse for superposition of the subsequent half cycle and reflected former half cycle (Fig. 1, inset). A nanotip is fabricated in the central interactive area to further enhance the field and induce electron emission. The emission charge is measured by a Keysight B2985A electrometer connected to the Faraday cup located at the gun exit. A temperature-controlled heating block is connected to the back of the electron gun to investigate the temperature-dependent field emission characteristics.

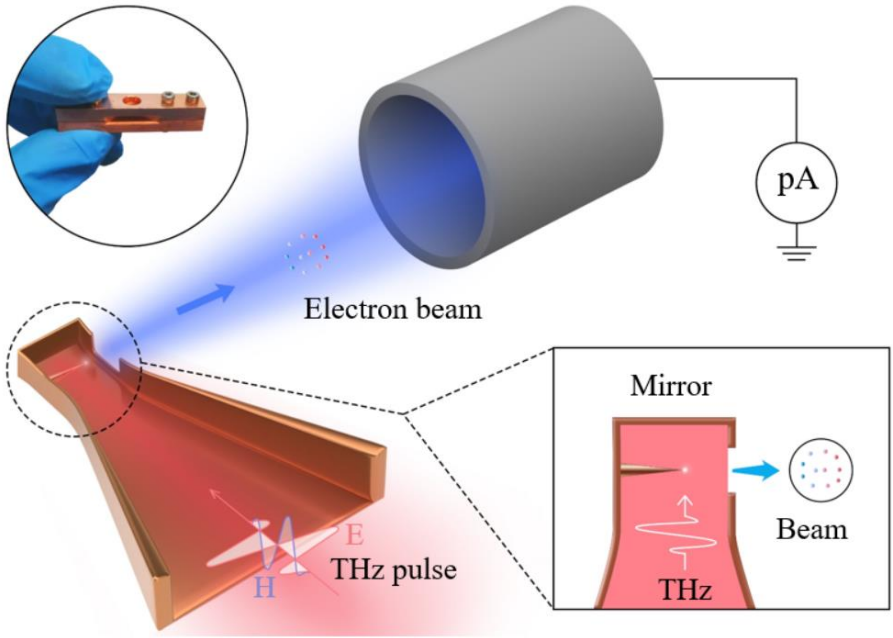

**Fig. 1 Experimental setup schematic.** The THz pulse generated by optical rectification in a lithium niobate crystal is coupled into the THz gun by an off-axis parabolic (OAP) mirror. The THz electric field ($E$) aligns with the electron propagation direction, and the magnetic field ($H$) orients perpendicularly to it. A Faraday cup is located at the gun exit to measure the emitted beam charge. The electrons are induced and accelerated by the extremely intense THz electric field at the tip. Inset: Enlarged view of the cross-section of the THz gun and a schematic of beam acceleration.

**Experimental observations and key problems**

The experimental results are presented in Fig. 2. The measured relationship between the emitted charge and THz energy shown in Fig. 2(a) displays a convex trend. This phenomenon considerably differs from the concave relationship predicted by the traditional S–N formula[30] $J=A/t_F^2 E^2 \exp(-v_F B/E)$. Here, $J$ represents the field emission current density; $E$ denotes the electrical field strength; *A* and *B* are constants that depend on the material's work function and surface characteristics; and $t_F$ and $v_F$ are special elliptic functions for the correction of the image charge effect. When the injected THz energy is increased to 3 µJ, the simulated corresponding electric field intensity reaches 16 GV·m$^{-1}$ at the cathode tip, and the detected emission charge ($Q_{emission}$) reaches 87 fC. Beyond this energy threshold, higher THz energies yield no substantial increase in the collected charge. In addition, contrary to conventional theoretical expectations that thermal excitation can enhance electron emission by increasing free-electron kinetic energy[30], temperature-dependent measurements demonstrate an inverse correlation between the cathode temperature and emission efficiency. The saturated $Q_{emission}$ decreases to 76 fC when the cathode temperature is increased from 295 K to 473 K. These counterintuitive phenomena challenge the application of existing thermal field emission theories in the ultrafast extremely intense field regime. To elucidate the cause, the microstructure evolution of the cathode tip was systematically analysed before and after the main experiment. The corresponding scanning electron microscopy (SEM) images of the tip are shown in Fig. 2(b). An ageing process was conducted to achieve geometric stabilization by applying THz pulses at room temperature (295 K) before the main experiment. Slight localized oxidation was observed on the lateral surface of the tip after the ageing process. Although field emission experiments conducted subsequently at emitter temperatures of 295 K and 473 K have led to progressive oxidation, no obvious oxidation has appeared at the apex from which the charge was emitted. The lack of obvious oxidation at the tip may be due to field evaporation during oxidation[27]. Furthermore, no obvious shape changes were observed in the apex. This observation suggests that the aforementioned saturation of emitted charge and its negative correlation with temperature arise from other factors, necessitating a more in-depth interpretation.

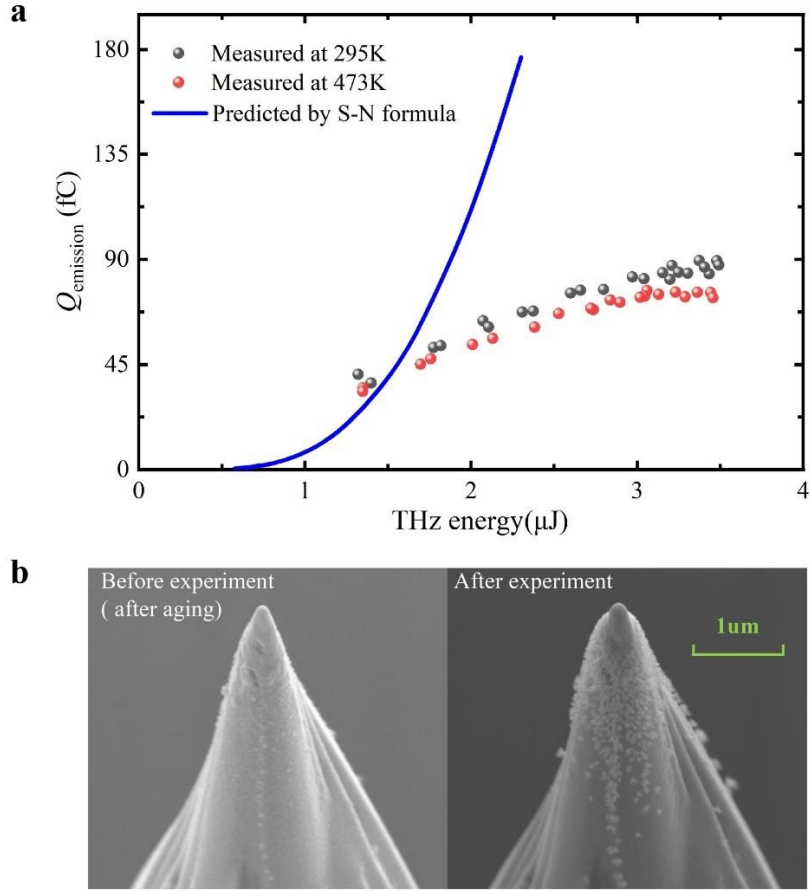

**Fig. 2 Experimental observations. a.** Relationship between the emitted charge and THz energy. The spheres represent measured data for emitter temperatures of 295 K (grey spheres) and 473 K (red spheres). The curve is calculated from the S–N formula, with the emission area and the peak electric fields on the tip under different THz energies determined via simulations. The peak field strength is 16 GV·m$^{-1}$ when the input THz energy reaches 3 μJ. **b.** SEM tip shape images before (after ageing) and after the formal experiment.

**Theory model and interpretation**

To interpret the observed experimental phenomena, we conducted a thorough study on the dynamic process of electron emission under ultrafast extremely intense THz fields. Indeed, the emitter's inherent constraints (finite density of SVEs and limited inner FER rate) prevent the infinite growth of field-emitted currents with increasing electric fields. Consequently, the simple scaling of the S–N formula to extremely high field strengths is physically untenable. A physical upper limit of the field emission current exists under specific conditions, providing a physically grounded interpretation of our experimental results. The maximum emission charge is analytically derived as the superposition of two components illustrated in Fig. 3: (1) the field emission current originating from SVEs and (2) the FER current generated by the diffusion of free electrons from the near-surface volume. The constraint of field emission currents originates fundamentally from charge conservation. Violating this equilibrium condition results in unphysical breaks in charge continuity. The field emission current arising from SVEs is derived separately in Part A, whereas the FER current inside the emitter is discussed separately in Part B. Integrating these results, Part C evaluates the coupled mechanisms and combined effects of these two currents, which are not obtained by simply adding their individual results.

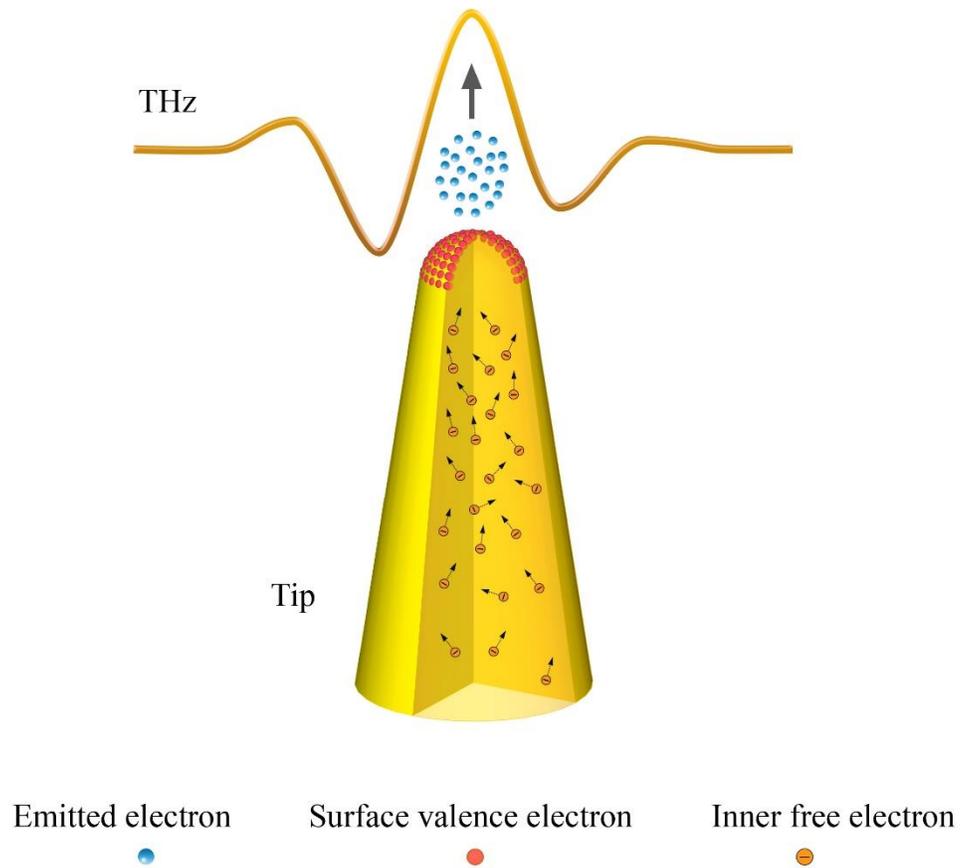

**Fig. 3 Illustration of the composition of field-emitted charge.** The field emission currents originate from surface valence electrons (SVEs). The currents arising from free-electron diffusion in the near-surface volume replenish the emitted surface valence electrons.

*Field emission current originating from SVEs*

The number of SVEs that can be emitted per surface copper atom under a sufficiently strong electric field is typically restricted to one. We disregard the possibility of two or more electrons tunnelling from a single atom, as the second ionization energy is typically much higher than the first ionization energy. Consequently, the charge density of SVEs ($\sigma_{SVE}$) is calculated as the total charge of SVEs per unit area and is limited by copper's atomic density ($n_{Cu} = 8.49 \times 10^{22}$ cm$^{-3}$). The initial $\sigma_{SVE}$ is $en_{Cu}^{2/3} = 3.09 \times 10^{-4}$ C·cm$^{-2}$ (see Method). Once surface charge depletion occurs without considering internal electron replenishment, no more surface electrons are emitted. The coefficients $A$ and $B$ in the S–N formula are governed by the intrinsic electronic properties and surface characteristics of the cathode made from copper. Since this formula fundamentally depends on the tunnelling probability and the initial population of tunnelling electrons, the coefficient $A$ necessitates proportional adjustment during dynamic emission processes due to the progressive depletion of surface charge density. Fig. 4(b) presents the current density ($J_{SVE}$) values calculated via the S–N formula with and without proportional adjustment of $A$, and the THz half-cycle signal (Fig. 4(a)) with a peak field strength of 16 GV/m is applied. The nonphysical current discontinuity arises in the results from the unmodified S–N formula, whereas the current exhibits continuous temporal evolution with instant proportional adjustment of $A$. This result indicates that the modified model provides a more reasonable physical description. The emission area is critical in field emission for evaluating the emission charge and comparing it with experimental data. The emission area in our electron gun is $1.3 \times 10^4$ nm² with a tip radius of 0.12 μm, which is determined through particle-in-cell (PIC) simulations in the CST Studio Suite[35] (see Method). The results of the emission charges ($Q_{emission}$) under different THz peak fields are shown in Fig. 4(c). The numerical results reveal saturation behaviour with an upper limit of ~40 fC in high-field regimes. While achieving essentially identical saturation compared with the traditional S–N formula, the modified model results in a smoother saturation curve. The underestimation of the experimental data stems from neglecting FER currents, which requires further consideration of inner free-electron diffusion and will be addressed in the following section.

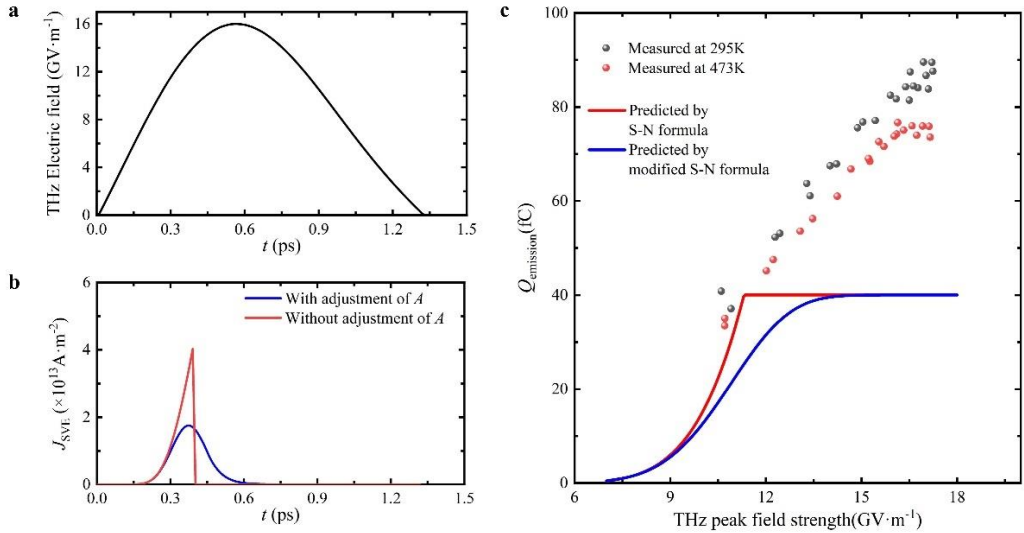

**Fig. 4 Field emission current originating from surface valence electrons without considering the inner replenishment current. a.** THz half-cycle time signal on the tip exported from the CST simulation results, with the experimentally measured THz signal serving as the input excitation. It is used to calculate the results in (**b**) and (**c**) (field strength scaled). **b.** Emission current densities calculated via the S–N formula with (blue curve, modified) and without (red curve) proportional adjustment of $A$. **c.** Calculated emission charges under different THz peak fields with the normalized signal (**a**) and the experimental measurements. The blue curve and red curve are the results calculated via the S–N formula with and without proportional adjustment of $A$, respectively. The grey spheres and red spheres represent the experimental results at cathode temperatures of 295 K and 473 K, respectively.

*Replenishment current due to free electron diffusion inside copper*

Owing to the extremely high collision frequency, the drift velocities of free electrons are orders of magnitude lower than their random moving velocities (comparable to the Fermi velocity $v_f$). Thus, the bulk-to-surface replenishment current originates predominantly from free-electron diffusion, with drift-induced currents being negligible. We therefore adopt the concept of equating the FER current with the diffusion current in subsequent treatment. The diffusion equation, a fundamental transport equation in physics and chemistry, is employed to characterize the distribution of the replenishment current over time and space. The diffusion equation is typically written as $\partial c/\partial t = D \partial^2 c/\partial t^2$. Here, $c$ denotes the concentration of the substance, which is the free-electron density ($n_{FE}$) in the calculation of free-electron diffusion; $D$ is the diffusion coefficient and is approximately inversely proportional to the absolute temperature. As established in Refs 36-38, metallic free-electron densities remain orders of magnitude lower than atomic densities. Only a few conduction electrons with energies close to the Fermi energy ($E_F$) exhibit stochastic mobility at characteristic Fermi velocities. From our experimental data, we determined the ratio of the free-electron density to the atomic density to be $2.21 \times 10^{-3}$. The determination of key physical quantities and the numerical finite difference method used to solve the diffusion equation are detailed in the Methods section.

The SVE density changes are neglected in this part. The calculated results obtained via the THz signal in Fig. 4(a) are shown in Fig. 5. In region 1, the inner FER current is large enough to support the field emission current. Therefore, the emission current $J_{FER}$ equals the result derived from the S–N formula. As the THz field increases and the emission proceeds, $n_{FE}$ starts to decrease near the surface. Since the diffusion coefficient $D$ is inversely proportional to temperature, in region 1, $n_{FE}$ decreases faster on the surface at higher temperatures because of the lower inner FER current, whereas $n_{FE}$ decreases more slowly at the inner position at higher temperatures. In region 2, the inner FER current is lower than the result from the S–N formula. $n_{FE}$ decreases to zero on the surface boundary. Since SVE density changes are neglected in this part, the emission current $J_{FER}$ is then equal to the inner FER current, which is strictly determined by the diffusion equation. As $n_{FE}$ decreases at the inner position during this process, the inner FER current decreases gradually. In region 3, as the THz field strength decreases, the inner FER current becomes larger than the S–N formula prediction again. $J_{FER}$ is equal to the calculated result from the S–N formula, and $n_{FE}$ increases simultaneously. As shown in Fig. 5(b), the inner FER current density is typically above $10^{12}$ A/m². This value is on the same order of magnitude as $e \cdot n_{FE0} \cdot v_f/6$, where $n_{FE0} \cdot v_f$ represents the free-electron particle flux and the denominator 6 accounts for the random motion in three-dimensional space. Before the emission current density surpasses the inner FER current density, the phenomenon of field emission saturation is not observed in the metal, indicating the necessity of extremely high local field strength. In addition, such a high emission current density could lead to melting of the emitter and result in explosive electron emission following plasma generation, thereby precluding the observation of saturation phenomena. Thus, a sufficiently short pulse duration is also necessary for field emission saturation. Ultrafast intense THz pulses are a good choice.

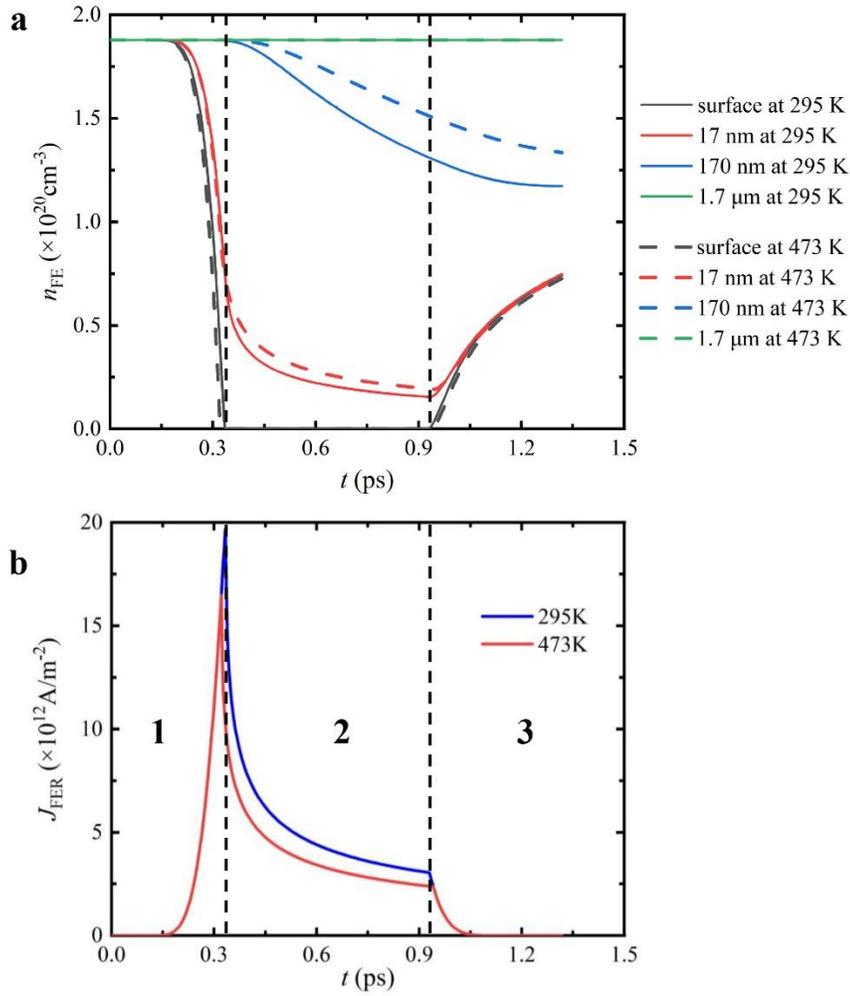

**Fig. 5 Calculated results of the $n_{FE}$ change (a) and real-time emission current $J_{FER}$ (b) using the THz signal in Fig. 4(a). a.** $n_{FE}$ at different depths (surface represented by the black curve, 0.14 μm by the red curve, 0.42 μm by the blue curve, and 2.1 μm by the green curve) from the surface at two temperatures (solid curves at 295 K and dashed curves at 473 K). **b.** Real-time emission current $J_{FER}$ (blue curve at 295 K and red curve at 473 K) neglecting surface charge density changes. Thus, in region 2, where the inner FER current is lower than the result from the S–N formula, the emission current is equal to the inner FER current to preserve a constant surface charge density.

*Combined effects*

In this section, we synthetically consider the total emitted charge contributed by the field emission current originating from the SVEs and the inner FER current due to free-electron diffusion. The calculated real-time emission current density considering both contributing mechanisms (see Method) is presented in Fig. 6(a). In region 1, the emission current density aligns with the prediction of the S–N formula, as sufficient free-electron diffusion maintains full charge compensation to the emitted SVEs. As the applied electric field increases, in region 2, the FER current becomes insufficient to fully compensate for the field emission current ($J_{\text{field}}$), leading to a progressive reduction in the surface charge density of the SVEs ($\sigma_{\text{SVE}}$). The actual $J_{\text{field}}$ is then calculated from the modified S–N formula. Compared with the results from the traditional S–N formula without these two constraints, this emission model reaches the peak earlier under a strong THz field and subsequently decreases more steeply because $\sigma_{\text{SVE}}$ decreases sharply, which benefits the generation of ultrafast bunches. At the end of this region, $J_{\text{field}}$ is equal to the FER current. In region 3, the FER current becomes sufficient to support $J_{\text{field}}$ calculated from the modified S–N formula, and the remaining FER current replenishes the SVEs. Although $\sigma_{\text{SVE}}$ increases, $J_{\text{field}}$ decreases because the dominant factor is the diminishing THz field.

The results of total emission charge ($Q_{\text{emission}}$) under different THz field strengths are shown in Fig. 6(b). If we assume that the field distribution at the tip is spatially uniform, the $Q_{\text{emission}}$ is calculated by integrating the product of the real-time emission current density and the emission area. The trend shown by the result (green curve in Fig. 6(b)) is consistent with the experimental data, despite discrepancies. Considering the actual spatially nonuniform field in the emission area calculated by CST (Fig. 6(b), inset), we performed numerical integration of the discrete charge contributions from individual surface elements. The result is plotted against the strength of the spatially nonuniform field as the red curve in Fig. 6(b). For direct comparison with the spatially uniform field result, the horizontal coordinate represents the average field strength over the emission area. This result is more consistent with the experimental data. The experimental and simulated results of the emission charge under different field strengths with different cathode temperatures are shown in Fig. 6(c). The emission charge saturation and the inverse relationship between the cathode temperature and saturated emission charge are successfully predicted by the model. In addition, the increase in the influence of the cathode temperature on the emission current with increasing THz electric field strength is also replicated by our numerical calculations.

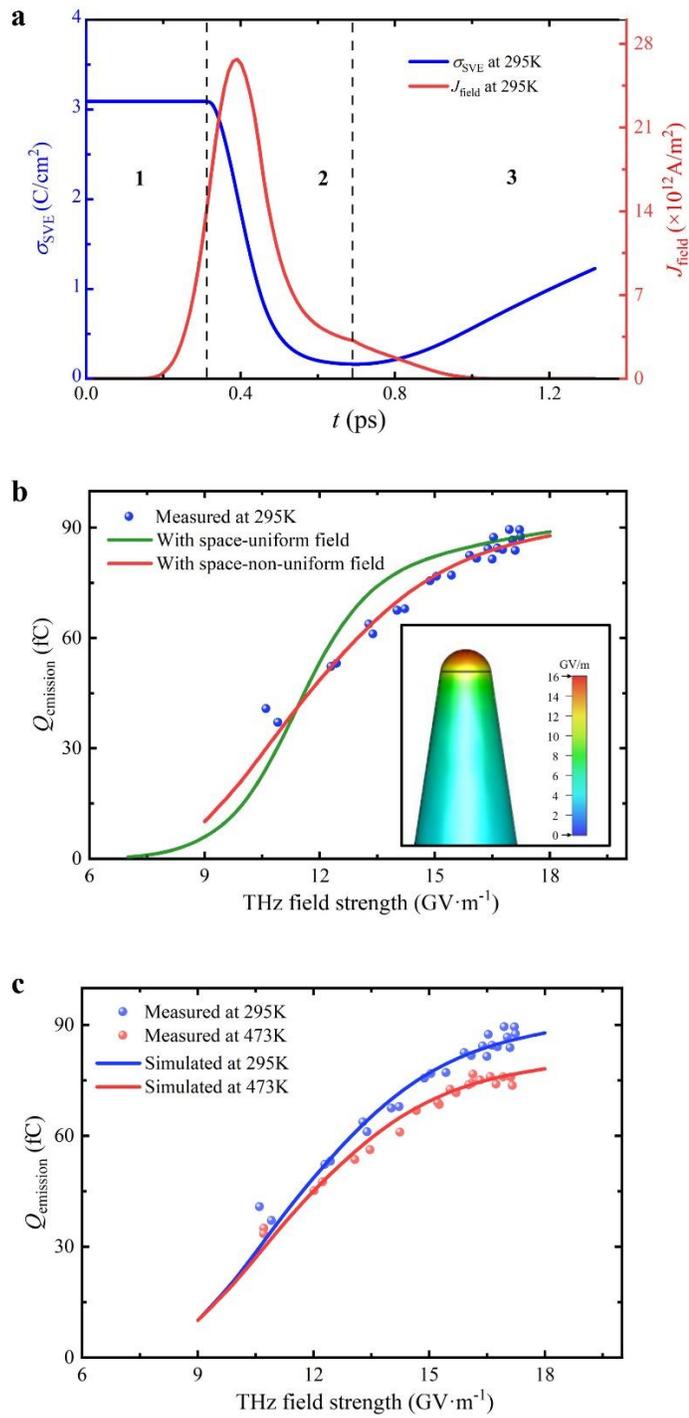

**Fig. 6 Results of considering surface charge variation and the FER current. a.** Real-time evolution of the surface charge density and emission current under the THz signal in Fig. 4(**a**). **b.**

Total emission charge under different THz field strengths. The experimental measurements are plotted as blue spheres. The green curve represents the results when a uniform field distribution over the emission area is assumed. The red curve shows the results when the nonuniform field distribution over the emission area is considered, with its horizontal coordinate representing the average field strength. Inset: simulated field distribution on the tip. **c.** Experimental and simulated results of the emission charge under different field strengths. The blue and red curves represent the theoretical predictions at cathode temperatures of 295 K and 473 K, respectively. The blue and red spheres represent experimentally measured data obtained at the corresponding cathode temperatures.

**Conclusions and outlook**

This work analyses the physical mechanism underlying the emission charge saturation from the metal tip in a THz-driven field emission electron gun. Our experiments reveal that at extremely intense field regimes exceeding 16 GV/m, charge saturation is fundamentally constrained by the total charge of surface valence electrons and the bulk-to-surface electron replenishment during picosecond-scale emission windows. In addition, the reduced saturated emitted charge caused by the temperature increase is well interpreted by our model. This model predicts unique characteristics of electron field emission under extremely high ultrafast electric field strengths, thereby enriching our understanding of the physical mechanisms underlying field emission. Moreover, it offers fresh insights into generating ultrafast electrons through the application of ultrashort high-intensity pulses to emit all SVEs within an extremely short period. The resulting electron beams are theoretically predicted to exhibit shorter pulse durations and higher current densities, providing advanced probes for ultrafast processes in matter. This model helps users select an appropriate field strength and pulse width for such pulses. As research continues to delve deeper into fundamental mechanisms and technological innovation progresses, ultrafast field emission electron source technology has emerged as an indispensable tool for investigating ultrafast dynamic processes. This advanced methodology, which offers unprecedented temporal resolution and precision, will undoubtedly catalyse groundbreaking discoveries across multiple scientific disciplines. We anticipate that its expanding applications will not only enhance our understanding of matter's transient states but also pave the way for transformative breakthroughs that considerably propel scientific discovery.

# References

1. Yang, Q. & Meng, S. Light-induced complete reversal of ferroelectric polarization in sliding ferroelectrics. *Phys. Rev. Lett.* 133, 136902 (2024).

2. Chang, Y. P. et al. Electronic dynamics created at conical intersections and its dephasing in aqueous solution. *Nat. Phys.* 21, 137–145 (2025).

3. Danz, T., Domrose, T. & Ropers, C. Ultrafast nanoimaging of the order parameter in a structural phase transition. *Science* 371, 371–374 (2021).

4. Chen, Y. et al. Observation of an anisotropic ultrafast spin relaxation process in large-area WTe2 films. *J. Appl. Phys.* 131, 163903 (2022).

5. Hennecke, M. et al. Ultrafast opto-magnetic effects in the extreme ultraviolet spectral range. *Commun. Phys.* 7, 191 (2024).

6. Loh, Z.-H. Studies of ultrafast molecular dynamics by femtosecond extreme ultraviolet absorption spectroscopy. *Chem. Lett.* 50, 965–973 (2021).

7. Mo, M. et al. Visualization of ultrafast melting initiated from radiation-driven defects in solids. *Sci. Adv.* 5, eaaw0392 (2019).

8. Mo, M. et al. Direct observation of strong momentum-dependent electron-phonon coupling in a metal. *Sci. Adv.* 10, eadk9051 (2024).

9. Ekeberg, T. et al. Observation of a single protein by ultrafast X-ray diffraction. *Light Sci. Appl.* 13, 15 (2024).

10. Buzzi, M. et al. Single-shot monitoring of ultrafast processes via x-ray streaking at a free electron laser. *Sci. Rep.* 7, 7253 (2017).

11. Kurtz, F. et al. Non-thermal phonon dynamics and a quenched exciton condensate probed by surface-sensitive electron diffraction. *Nat. Mater.* 23, 890–897 (2024).

12. Wang, X., Musumeci, P., Lessner, E. & Goldstein, J. *Report of the Basic Energy Sciences Workshop on the Future of Electron Sources, September 8-9, 2016* (2016). https://www.osti.gov/servlets/purl/1616511.

13. Krüger, M. & Hommelhoff, P. *Structural Dynamics with X-Ray and Electron Scattering* (Royal Society of Chemistry, 2023).

14. Filippetto, D. et al. Ultrafast electron diffraction: Visualizing dynamic states of matter. *Rev. Mod. Phys.* 94, 045004 (2022).

15. Leedle, K. J. et al. High gradient silicon carbide immersion lens ultrafast electron sources. *J. Appl. Phys.* 131, 134501 (2022).

16. Hoffrogge, J. et al. Tip-based source of femtosecond electron pulses at 30 keV. *J. Appl. Phys.* 115, 094506 (2014).

17. Ehberger, D. et al. Highly coherent electron beam from a laser-triggered tungsten needle tip. *Phys. Rev. Lett.* 114, 227601 (2015).

18. Kim, H. Y. et al. Attosecond field emission. *Nature* 613, 662–666 (2023).

19. Yanagisawa, H. et al. Delayed electron emission in strong-field driven tunnelling from a metallic nanotip in the multi-electron regime. *Sci. Rep.* 6, 35877 (2016).

20. Dienstbier, P. et al. Tracing attosecond electron emission from a nanometric metal tip. *Nature* 616, 702–706 (2023).

21. Garg, M. et al. Real-space subfemtosecond imaging of quantum electronic coherences in

molecules. *Nat. Photonics* 16, 196–202 (2021).

22. Bormann, R., Gulde, M., Weismann, A., Yalunin, S. V. & Ropers, C. Tip-enhanced strong-field photoemission. *Phys. Rev. Lett.* 105, 147601 (2010).

23. Wimmer, L. et al. Terahertz control of nanotip photoemission. *Nat. Phys.* 10, 432–436 (2014).

24. Vella, A. et al. High-resolution terahertz-driven atom probe tomography. *Sci. Adv.* 7 (2021).

25. Li, S. et al. Subcycle surface electron emission driven by strong-field terahertz waveforms. *Nat. Commun.* 14, 6596 (2023).

26. Colmey, B., Paulino, R. T., Beaufort, G. & Cooke, D. G. Sub-cycle nanotip field emission of electrons driven by air plasma generated THz pulses. *Appl. Phys. Lett.* 126 (2025). doi: 10.1063/5.0238527

27. Matte, D. et al. Extreme lightwave electron field emission from a nanotip. *Phys. Rev. Res.* 3 (2021). doi: 10.1103/PhysRevResearch.3.013137

28. Li, S. & Jones, R. R. High-energy electron emission from metallic nano-tips driven by intense single-cycle terahertz pulses. *Nat. Commun.* 7, 13405 (2016).

29. Fowler, R. H. & L. Nordheim. Electron emission in intense electric fields. *Proc. R. Soc. A* 119, 173–181 (1997).

30. Murphy, E. L. & Good, R. H. Thermionic emission, field emission, and the transition region. *Phys. Rev.* 102, 1464–1473 (1956).

31. Tan, N. et al. Mechanism analysis of field electron emission of titanium. *Phys. Scr.* 98,

045005 (2023).

32. Chan, W. J., Ang, Y. S. & Ang, L. K. Thermal-field electron emission from three-dimensional Dirac and Weyl semimetals. *Phys. Rev. B* 104, 245420 (2021).

33. Keldysh, L. V. Ionization in the field of a strong electromagnetic wave. *Rev. Mod. Phys.* 94, 045004 (2022).

34. Shlomo, N. & Frumker, E. In situ characterization of laser-induced strong field ionization phenomena. *Light Sci. Appl.* 14, 166 (2025).

35. CST Studio Suite. (2021). https://www.3ds.com/products/simulia/cst-studio-suite.

36. Ibach, H. & Lüth, H. *Solid-State Physics: An Introduction to Principles of Materials Science* (Springer, 1995).

37. Palenskis, V. & Žitkevičius, E. Summary of new insight into electron transport in metals. *Crystals* 11, 622 (2021).

38. Lang, P. F. Fermi energy, metals and the drift velocity of electrons. *Chem. Phys. Lett.* 770, 138447 (2021).

**Online methods**

*THz generation and transmission*

The experimental setup for THz generation is shown in Extended Data Fig. 1. The 800 nm near-infrared laser pulse with a width of ~400 fs is divided through a beam splitter (T/R = 90/10). The reflected light serves as a probe for THz time-domain signal measurement. The transmitted light is shaped into a tilted wavefront via a diffraction grating. The quasi-single-cycle THz pulse is generated by tilted wavefront light through optical rectification in a congruent lithium niobate (cLN) crystal. The diffraction grating, two cylindrical lenses, and cLN crystal are installed on a multidimensional mirror frame to optimize the THz conversion efficiency.

The generated THz wave initially passes through two assembled gold-coated mirrors, which regulate its propagation height and directional alignment while changing the THz polarization from a vertical to horizontal orientation for compatibility with the subsequent horizontal beamline. Afterwards, the THz wave propagates through two gold-coated off-axis parabolic mirrors (OAP 1 and OAP 2) to create a parallel THz beam for efficient long-distance propagation. The OAP with a hole (OAP 3) is mounted on a translation stage for THz measurement when it is in the transmission path. The measurement employs either a ZnTe crystal paired with a balanced photodiode for THz time-domain signal acquisition or a Golay cell for THz energy measurement. The measured THz signal is shown in the lower right corner of Extended Data Fig. 1. Finally, the THz wave passes through the CF63 TPX window and enters the vacuum chamber. A dedicated OAP (OAP 4) creates a THz focus (inset in Extended Data Fig.

1) for efficient coupling of THz energy into the electron gun, enabling subsequent field emission and electron acceleration.

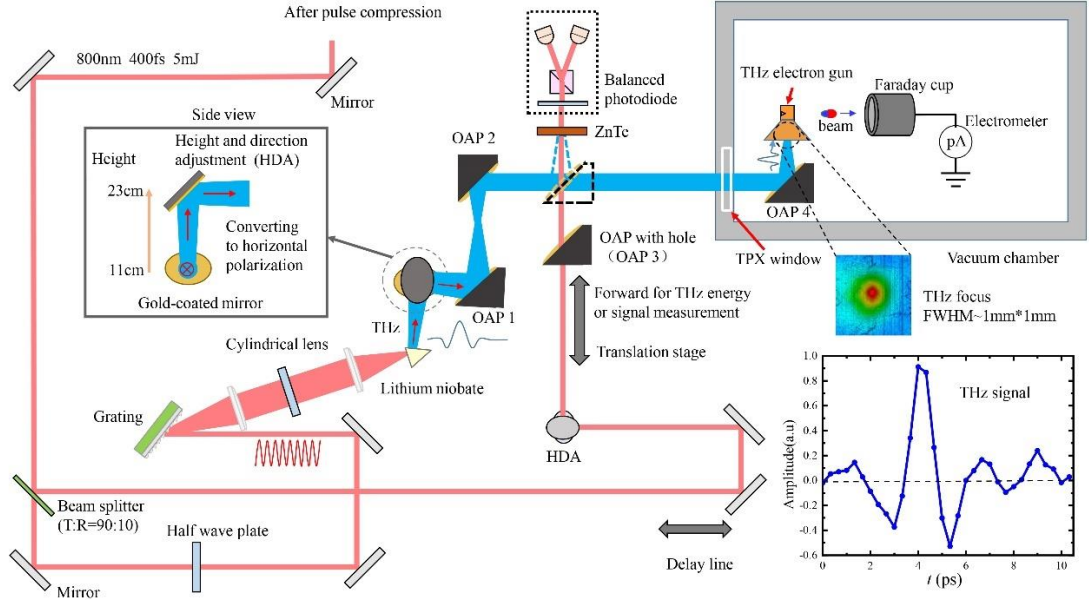

**Extended data Fig. 1** | THz-driven field emission experimental setup

*Electron gun structure*

The THz-driven electron gun shown in Extended Data Fig. 2 is made of oxygen-free copper and features a tapered rectangular waveguide horn structure[38] with a nanotip emitter in the centre. The gun structure is meticulously designed and optimized to achieve higher field enhancement and electron acceleration. The tapered section maintains a constant characteristic impedance ($b(z) = b_0\sqrt{1 + (\frac{z}{z_{rb}})^2}$ , $\frac{b(z)}{\sqrt{a(z)^2 - (\frac{\lambda}{2})^2}} = \frac{b_0}{\sqrt{a_0^2 - (\frac{\lambda}{2})^2}}$) to minimize wave reflection. Simultaneously, the waveguide entrance is nearly impedance-matched to the vacuum wave impedance. This design not only enhances the coupling efficiency of THz energy from free space into the rectangular waveguide but also intensifies the field strength within the gun's interaction region. The uniform rectangular section is terminated with a short-circuit boundary condition acting as a mirror. Through precise optimization of the cavity length, phase-synchronized superposition of incident and reflected waves generates constructive interference, further strengthening the effective accelerating field gradient. Furthermore, the nanotip prepared by plasma focused ion beam (PFIB) etching further amplifies the field, building upon the field enhancement provided by the preceding waveguide structure, resulting in a multiplicative increase in field strength. SEM characterization of the total cathode morphology after the ageing process at room temperature is displayed in the lower left corner of Extended Data Fig. 2. This architecture amplifies the local electric field intensity by approximately 100 times according to electromagnetic simulations via CST Studio Suite software. The optimized structure parameters are summarized in Extended Data Table 1. Operating in $TE_{10}$ mode, this configuration ensures axial alignment between the THz pulse electric field vector and the electron propagation direction.

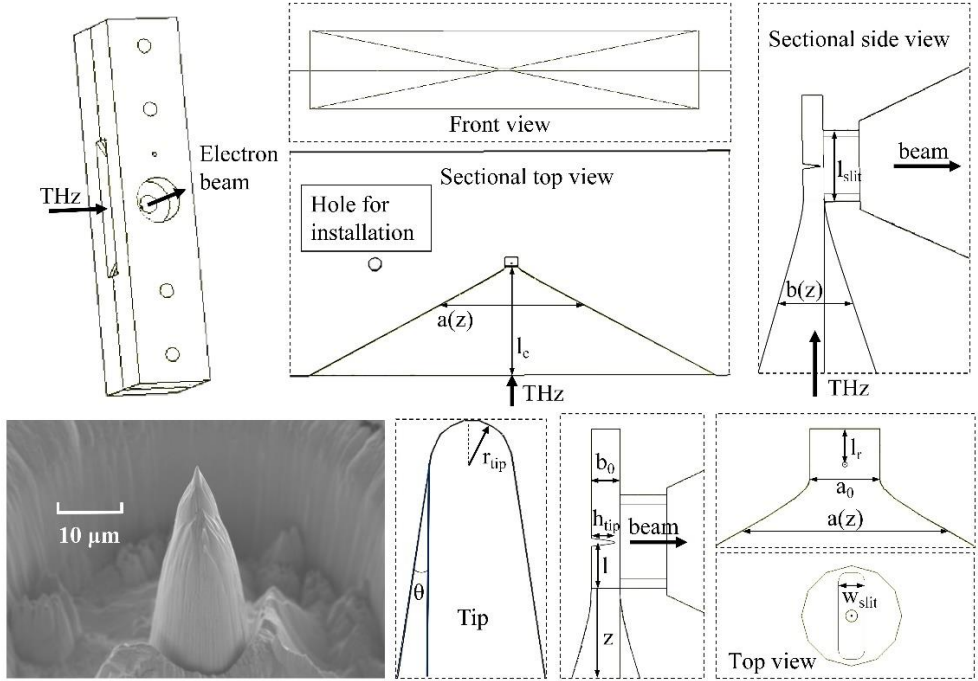

**Extended data Fig. 2 |** Structure of a THz-driven electron gun

**Extended Data TABLE I.** Optimized structural parameters of the THz electron gun

| Parameter | Value | Description |
|---|---|---|
| $a_0$ | 0.47 mm | Uniform waveguide, wide side |
| $b_0$ | 60 μm | Uniform waveguide, narrow side |
| $Z_{rb}$ | 88.6 μm | Shape factor of the coupler |
| $l_c$ | 4 mm | Coupler length |
| $l$ | 0.1 mm | Initial length of the uniform waveguide |
| $l_r$ | 0.24 mm | Reflective cavity length |
| $l_{slit}$ | 0.2 mm | Length of the slit beam exit |
| $w_{slit}$ | 60 μm | Width of the slit beam exit |
| $h_{tip}$ | ~50 μm | Height of the tip emitter |
| $\theta$ | ~8.5° | Cone angle of the emitter |
| $r_{tip}$ | ~120 nm | Tip radius of the emitter |

*Emission area determination*

The emitted charge or the total emission current is usually a directly measurable physical quantity. To establish a connection with the current density resulting from field emission calculations, the emission area is crucial. For a large emission surface, the specific emission area is often difficult to accurately determine because of the microstructural field enhancement caused by surface irregularities. Using a nanotip emitter allows the emission area to be precisely confined to the apex. A more reliable emission area can be obtained by precisely measuring the tip morphology via SEM. For our electron gun structure, the acceptance of the gun's anode aperture is an additional factor that needs to be considered. To precisely determine the corresponding emission area, we performed PIC simulations in the CST Studio Suite. The typical results are presented in Extended Data Fig. 3. Every emission particle has a unique identification number in the exported information, including positions, moments and charges. According to the identification numbers of electrons that travel through the anode slit, one can easily obtain their initial emission areas. Consequently, the total effective emission area in our experiment is precisely determined to be $1.3 \times 10^4$ nm$^2$.

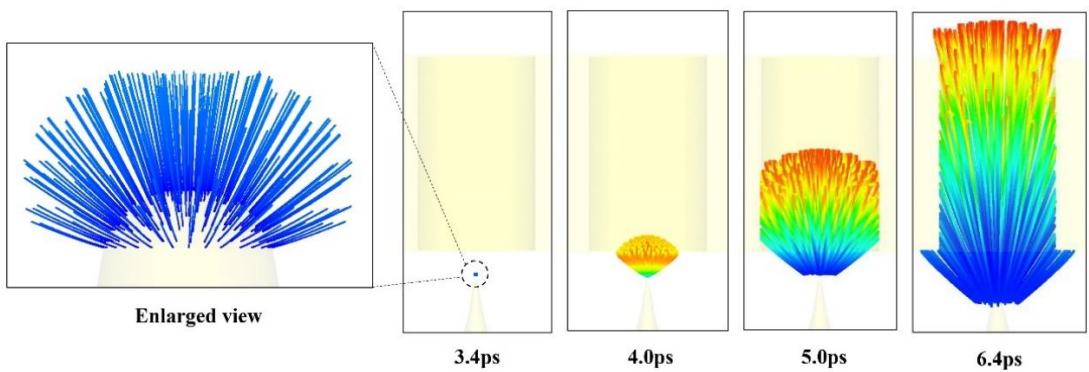

**Extended data Fig. 3 |** Results of PIC simulation in the CST studio suite

*Key physical quantity determination*

The atomic density of oxygen-free copper is calculated via the equation $n_{Cu}=\rho N_A/M_{mol}=8.49\times10^{22}$ cm$^{-3}$, where $\rho$ indicates the density of copper ($\rho=8.96$ g·m$^{-3}$); $M_{mol}$ denotes its molar mass ($M_{mol}=63.55$ g·mol$^{-1}$); and $N_A$ represents Avogadro's number ($N_A=6.02\times10^{23}$ mol$^{-1}$). The copper surface number density is derived as the number of copper atoms in a single layer of copper per unit area, which is $n_{Cu}\times$layer thickness. The layer thickness is equal to $n_{Cu}^{-1/3}$. Thus, the copper surface number density is $n_{Cu}^{2/3}$. Since the number of SVEs that can be emitted per surface copper atom is restricted to one, the SVE charge density is $en_{Cu}^{2/3}=3.09\times10^{-4}$ C·cm$^{-2}$.

The physical quantities critical for solving the electron diffusion equation cannot be directly measured experimentally. Typically, these quantities are inferred by relating them to measurable macroscopic properties via theoretical models, allowing their determination through the fitting of experimental data. According to the analyses in 36-38, metallic free-electron densities remain orders of magnitude lower than atomic densities. Only a few part of conduction electrons with energies close to the $E_F$ exhibit stochastic mobility at characteristic Fermi velocities. However, the ratio of the free-electron density to the atomic density is difficult to determine accurately. The results from different theoretical analyses may vary by up to two orders of magnitude, from 10$^{-2}$ to 10$^{-4}$. We adopt the ratio α and determine it by fitting our experimental data. The assumption that the free-electron density is almost independent of temperature[36,38,40] is accepted here. The diffusion coefficient $D$ of the random moving particles with a mean free path of $\lambda$ and a mean velocity of $\bar{v}$ is $1/3\lambda\bar{v}$, which is easily found in standard physics textbooks. For the randomly moving free electrons inside the copper, the mean velocity is the Fermi velocity $v_f = (2E_F/m)^{1/2}$, and the mean free path $l_f$ is equal to $v_f \times \tau$. $\tau$ is the mean collision time or lifetime of free electrons in metals, which is $\hbar/kT$[40]. The calculated value of $\tau$ is $2.6 \times 10^{-14}$ s at 295 K, which is on the same order of magnitude as the analyses in 36-38. The Fermi energy $E_F$ of copper is 7.03 eV, corresponding to a Fermi energy velocity of $1.57 \times 10^6$ m·s$^{-1}$ [36]. Since $\tau$ is inversely proportional to the temperature, the diffusion coefficient $D$ is inversely proportional to the temperature. After these parameters are substituted, the diffusion coefficient $D$ for copper is calculated to be $2.13 \times 10^{-2}$ m$^2$·s$^{-1}$ at room temperature (295 K) and $1.33 \times 10^{-2}$ m$^2$·s$^{-1}$ at 473 K.

*Electron diffusion equation solving*

The diffusion equation is $\partial n_{FE}/\partial t = D \partial^2 n_{FE}/\partial t^2$. The numerical finite difference method is adopted to solve the diffusion equation. The difference equation is derived as follows:

$$\frac{n_{j\Delta x}^{(i+1)\Delta t} - n_{j\Delta x}^{i\Delta t}}{\Delta t} = D \frac{\left(n_{(j+1)\Delta x}^{i\Delta t} - n_{j\Delta x}^{i\Delta t}\right) - \left(n_{j\Delta x}^{i\Delta t} - n_{(j-1)\Delta x}^{i\Delta t}\right)}{\Delta x^2} \tag{1}$$

Equation (1) can be rewritten as the following iterative algebraic equation:

$$n_{j\Delta x}^{(i+1)\Delta t} = \frac{D\Delta t}{\Delta x^2}\left(n_{(j+1)\Delta x}^{i\Delta t} + n_{(j-1)\Delta x}^{i\Delta t} - 2n_{j\Delta x}^{i\Delta t}\right) + n_{j\Delta x}^{i\Delta t} \tag{2}$$

Here, $n_{j\Delta x}^{i\Delta t}$ represents the free-electron density at the spatial position of $j\Delta x$ and temporal coordinate of $i\Delta t$, where $\Delta x$ and $\Delta t$ represent the spatial and temporal discrete steps, respectively. To ensure numerical convergence, $\Delta x = 1.7$ nm and $\Delta t = 1.12 \times 10^{-4}$ ps are selected for our simulations. This selection guarantees that the increment of $n_{j\Delta x}^t$ within each time step remains within a moderate range < 10%, thereby preventing excessively large increases that could lead to instability or inaccuracy in the simulation. In this equation, the index $j < j_{max}=1500$ is needed because the most inner position is chosen deep enough to be not disturbed during the emission time determined by the THz field signal. $j > 0$ is also needed because the equation of boundary conditions differ from the aforementioned iterative equation. The free-electron density at the boundary is calculated via the emission current density acquired from the S–N formula (in part B, the change in surface charge is ignored). at each time step:

$$n_0^{i\Delta t} = \begin{cases} n_0^{i\Delta t} - \dfrac{J_{SN}^{i\Delta t}\Delta t}{e\Delta x}, & n_0^{i\Delta t} > \dfrac{J_{SN}^{i\Delta t}\Delta t}{e\Delta x} \\ 0, & n_0^{i\Delta t} \leq \dfrac{J_{SN}^{i\Delta t}\Delta t}{e\Delta x} \end{cases} \quad (3)$$

Here, $J_{SN}^{i\Delta t}$ represents the emission current density acquired from the S–N formula at the temporal coordinate of $i\Delta t$. The initial number density (at $i=0$) of the random-moving electrons is equal to $n_{EF0}$. After the boundary condition is calculated, the iterative algebraic equation (2) can be applied on the basis of this boundary condition. Because the emission current has been considered at the start of the time step, the result of electron diffusion within this time step should be charge conservative. Therefore, after all the fee-electron densities at all positions are calculated, the boundary condition needs to be updated to ensure the conservation of charge within the time step, and this updated boundary condition contributes to the emission current at the start of the next time step. Finally, the time step index $i < i_{max}$ is needed because the result at time $i_{max}\Delta t$ has been calculated in the iterative process at the time step index $i=i_{max}$-1. Using the above method, one can numerically calculate the change in the density of internal free electrons due to free electron diffusion.

*Combination of both mechanisms*

In this part, the fundamental calculation methods remain consistent with the method described above but with additional details and refinements related to the modified S–N formula. The modified S–N formula is used to calculate the boundary condition. In the initial stage of the half cycle, the THz field strength is small, and $n_0^{i\Delta t} > \frac{J_{SN}^{i\Delta t} \Delta t}{e\Delta x}$ is satisfied. All surface valence charge emission can be replenished by internal free-electron diffusion. Therefore, $\sigma_{SVE}$ remains unchanged, and the emission current density acquired from the modified S–N formula ($J_{SNm}^{i\Delta t}$) at the temporal coordinate of $i\Delta t$ equals to $J_{SN}^{i\Delta t}$. As the THz field increases with time, $n_0^{i\Delta t} > \frac{J_{SNm}^{i\Delta t} \Delta t}{e\Delta x}$ is not satisfied, and the surface charge density changes. In each time step, the reduction of $\sigma_{SVE}$ is calculated as the product of $\Delta t$ and the difference between the emission current density determined from the modified S–N formula and the FER current density. The real emission current density, which is the results from the modified S–N formula, will be lower than that of the traditional S–N formula. Near the end of the emission half cycle, the THz field strength decreases to a sufficiently small value such that $n_0^{i\Delta t} > \frac{J_{SNm}^{i\Delta t} \Delta t}{e\Delta x}$ is satisfied. In this stage, the FER current not only supports the emission current determined by the modified S–N formula, but also replenishes the surface charge. In each time step, the increase in $\sigma_{SVE}$ is calculated as the product of $\Delta t$ and the difference between the FER current density and the emission current density determined from the modified S–N formula.

The parameter α, which is fitted from our experimental data in this part, has a value of

2.21×10$^{-3}$, which is in the theoretical range of 10$^{-2}$ to 10$^{-4}$.

**References**

39. Fallahi, A. & Kärtner, F. Design strategies for single-cycle ultrafast electron guns. *J. Phys. B: At. Mol. Opt. Phys.* 51, 144001 (2018).

40. Devillers, M. A. C. Lifetime of electrons in metals at room temperature. *Solid State Commun.* 49, 1019–1022 (1984).

**Data availability**

The data that support the plots within this paper and other findings of this study are available from the corresponding author upon reasonable request.

## Acknowledgements

The authors acknowledge support from the National Natural Science Foundation of China (NSFC Grant No. 12035010 and No.12405179).